\documentclass{pasj00}

\begin{document}
\SetRunningHead{R. S. Furuya et al.}{A Giant Flare on T Tauri Star Observed 
at Millimeter Wavelengths}
\Received{2003/08/27}
\Accepted{2003/10/24}

\title{A Giant Flare on A T Tauri Star Observed at Millimeter Wavelengths}

%
 \author{%
   Ray S.~\textsc{Furuya}\altaffilmark{1}
  \thanks{Present Address~:~Division of Physics, 
Mathematics, and Astronomy, 
California Institute of Technology, MS 105-24, Pasadena, 
CA 91125, U.S.A., ~\texttt{rsf@astro.caltech.edu}},
   Hiroko \textsc{Shinnaga}\altaffilmark{2},
   Kouichiro \textsc{Nakanishi}\altaffilmark{3},
   Munetake \textsc{Momose}\altaffilmark{4},
   and
   Masao \textsc{Saito}\altaffilmark{1}
}
 \altaffiltext{1}{National Astronomical Observatory, Osawa 2-21-1, Mitaka, Tokyo 181-8588}
 \altaffiltext{2}{Harvard-Smithsonian Center for Astrophysics, Sub-Millimeter Array, 
P.O. Box 824, Hilo, HI 96721, U.S.A.}
 \altaffiltext{3}{Nobeyama Radio Observatory, Nobeyama, Minamimaki, Minamisaku, Nagano 384-1305}
 \altaffiltext{4}{Institute of Astrophysics and Planetary Sciences, Ibaraki University, 
Bunkyo 2-1-1, Mito, Ibaraki 310-8512}


\maketitle

\begin{abstract}
We have conducted multi-epoch synthesis imaging of $\lambda=$2 and 3 millimeter 
(mm) continuum emission and near infrared K band (2.2 $\mu$m) imaging of 
a flare event in January 2003 that occurred on the young stellar object
GMR-A which is suggested to be a weak-line T Tauri star in the Orion cluster.
Our mm data showed that the flare activity lasted at least over 13 days,
whereas the K-band magnitude did not change during this event.
In addition, we have succeeded in detecting short time
variations of flux on the time scales of 15 minutes.
The total energy of the flare is estimated to be $\sim 10^{35-36}$ erg,
which makes it one of the most energetic flares reported to date.
Comparing the mm continuum luminosities with reported X-ray luminosities,
we conclude that the mm flare was similar in nature to solar and other 
stellar flares.
Our results will be a crucial step toward understanding
magnetically induced stellar surface activities in T Tauri stars.
\end{abstract}

\section{Introduction}

Recent radio and X-ray observations have revealed that 
pre-main-sequence stars show flare activity as seen in the Sun and
late-type stars.
The radio emission has been attributed to gyrosynchrotron emission
from mildly relativistic electrons and is preferentially seen in
X-ray bright stars.~
Luminosities of these flares at centimeter (cm) and 
those in soft X-rays are known to correlate well with each other
(G\"udel 2002; references therein).
Moreover, \citet{feld95} discovered a correlation 
between the peak temperature of a flare ($T$) and its volume emission 
measure ($EM$) derived from X-ray spectra
of solar and stellar flares.
Shibata \& Yokoyama (1999, 2002 hereafter SY02) attempted to explain 
the $EM-T$ relation by magnetohydrodynamic models.\par

Stellar X-ray emission is highly time variable, but variability 
at millimeter (mm) wavelengths is poorly understood.
This Letter\footnote{Based on the results from the 
Nobeyama Radio Observatory and Okayama Astrophysical Observatory
which are branches of the National Astronomical Observatory, operated by the
Ministry of Education, Culture, Sports, Science and Technology of Japan.}, 
presents in detail our mm interferometric observations of a flare that
occurred in 2003 January in GMR-A. 
Initial results were reported in Nakanishi et al. (2003).
The flare was originally discovered by Bower, Plambeck \& Bolatto (2003a) 
using the BIMA array on January 20 UT at 86 GHz and
subsequently reported by Bower et al. (2003b, hereafter BPB03).
GMR-A is identified as a member of the Orion cluster 
(M42; distance of 450 pc) from early 
VLA observations (Garay, Moran \& Reid 1987).
BPB03 identified that the object is a K5 type weak-line T Tauri star 
(WTTS; bolometric luminosity $\sim 6 L_{\odot}$) by
near-infrared (NIR) spectroscopy.
During the flux monitoring campaign at cm wavelengths over 7 months in 1990,
GMR-A showed one radio flare (Felli et al. 1993).

\section{Observations}

Synthesis imaging of the $\lambda=$2 and 3 mm 
continuum emission was carried out using the Nobeyama Millimeter Array 
(NMA) with the most expanded configuration (Table \ref{tbl:obs}).
Note that the two frequency bands were not observed simultaneously.
The phase tracking center was set at
R.A.$=$\timeform{5h35m11s.8},
Decl.$=$\timeform{-5D21'49''.2} (J2000)
and the field of view (FOV) was \timeform{70''} 
at 3 mm and \timeform{50''} at 2 mm.
For the backend, the 1 GHz bandwidth correlator was employed.
We used the continuum source 0420$-$014 as a bandpass calibrator and 
0528$+$134 as a phase and gain calibrator.
From observations of Uranus, we estimate the fluxes of 0528$+$134 to be
2.56$\pm$0.5~Jy at 98 GHz and
2.11$\pm$0.4~Jy at 147 GHz.

\begin{figure}
  \begin{center}
  \FigureFile(69mm,69mm){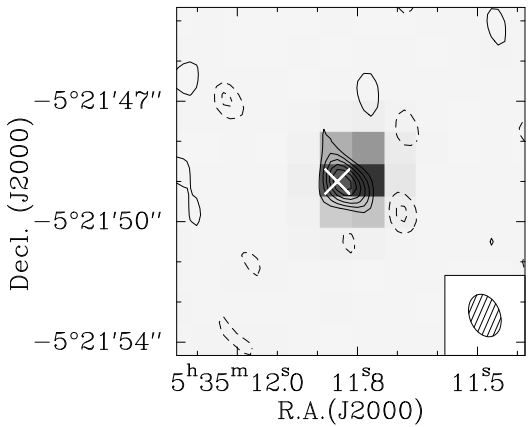} 
  \end{center}
  \caption{
An overlay of the NMA 2 mm continuum emission map (contour) 
taken on 2003 January 25 and the OASIS 2.2 $\mu$m image on January 18.
Solid and dashed contours indicate, respectively, positive and negative
ones starting at 2$\sigma$ level with $\pm 1\sigma$ step of
the noise level (Table \ref{tbl:obs}).
The cross indicates the peak position of the 86 GHz continuum 
emission (BPB03).
The synthesized beam size of the NMA observations is shown in the bottom
right corner. 
}
\label{fig:overlay}
\end{figure}

The NIR K-band ( $\lambda=$2.2~$\mu$m) data were taken with  
the Okayama Astrophysical System for Infrared imaging and Spectroscopy 
(OASIS) mounted on the 1.88 m telescope of the Okayama Astrophysical Observatory.
OASIS has a FOV of \timeform{4'} with an image pixel size of \timeform{0.''97}. 
The observations were done with seeing of 
$\lesssim$\timeform{1.7''} 
on 2003 January 18.451 and 21.459 UT and its calibration was 
performed using 23 stars reported in Muench et al. (2002).

\begin{table*}[ht]
\begin{center}
\caption{Summary of the NMA Observations toward the GMR-A Radio Flare in 2003 January}
\label{tbl:obs}
\begin{tabular}{lrrcrcc}
\hline 
UT 2003   & \multicolumn{2}{c}{Time Range of HA\footnotemark[$*$]} & Time\footnotemark[$\dagger$] & Frequency & 
Sensitivity\footnotemark[$\ddagger$] & $\theta_{\rm maj}\times\theta_{\rm min}$\footnotemark[$\S$] \\
          & \multicolumn{2}{c}{(hh:mm:ss.s)} & (hrs)       & (GHz)     & (mJy beam$^{-1}$) & (arcsec$\times$arcsec)  \\
\hline 
Jan.24.56 &    02:02:29.9 &     03:15:37.9 &  1.00  &  97.782 &  4.4 & 2.51$\times$1.18 \\
Jan.25.55 &    00:31:08.8 &     01:45:22.1 &  1.00  & 146.969 &  5.1 & 1.32$\times$0.91 \\
Jan.25.61 &    02:30:29.5 &     03:18:37.5 &  0.750 &  97.782 &  5.4 & 2.28$\times$1.21 \\
Jan.28.44 & $-$02:46:34.1 &  $-$00:46:13.7 &  1.90  &  98.243 &  4.1 & 2.25$\times$1.35 \\
Jan.28.58 &    00:42:59.0 &     03:02:22.6 &  1.75  & 141.359 &  2.4 & 1.65$\times$0.92 \\
Feb.02.57 &    00:32:38.8 &     03:22:05.5 &  2.25  &  99.023 &  4.1 & 2.89$\times$2.50 \\
Feb.07.57 &    01:08:22.9 &     03:41:48.3 &  2.00  &  99.023 &  2.8 & 2.58$\times$1.35 \\
Feb.15.37 & $-$04:02:00.0 &     00:07:37.6 &  3.25  &  99.023 &  3.1 & 2.85$\times$1.56 \\
\hline
\end{tabular}
\end{center}
 \footnotemark[$*$] Time ranges of the observations given with starting (left) and ending 
(right) hour angles (HAs),
 \footnotemark[$\dagger$] Effective integration times,
 \footnotemark[$\ddagger$] RMS image noise levels,
 \footnotemark[$\S$] Synthesized beam sizes along major ($\theta_{\rm maj}$) 
and minor ($\theta_{\rm min}$) axes.
\end{table*}

\section{Results and Discussion}
\subsection{Overall Properties:~ The Maps and Time Variations}
\label{ss:overall}

Fig.~\ref{fig:overlay} is a map of the continuum emission at 2 mm 
superposed on a 2.2 $\mu$m image. 
The peak positions of the 2 mm and 2.2 $\mu$m emission
coincide with each
other within errors ($\lesssim\timeform{1''}$), 
and their positions agree with the previous measurements with
BIMA and VLA (e.g., Felli et al 1993).
The K band flux was measured to be 
9.625$\pm$0.032 mag on January 18.451 and
9.586$\pm$0.047 mag on January 21.46.
There was no significant flux
change at K band during the most prominent mm flare 
on January 20.188 (BPB03).  
Furthermore, the above measurements are consistent with 
those in 1997 (Muench et al. 2002) and in 
1999 (Hillenbrand \& Carpenter 2000) within 0.05 mag,
suggesting that the K band flux is stable.
Comparing our K band data with those of H band
(Hillenbrand \& Carpenter 2000),
GMR-A is embedded in dust 
($\left (H-K\right )$~$=2.311\pm$0.047 mag).
Such strong infrared excess is more consistent with a
classical T Tauri star rather than a WTTS,
as BPB03 argued as well.

\begin{figure}
\centerline{\includegraphics[angle=0,width=74mm]{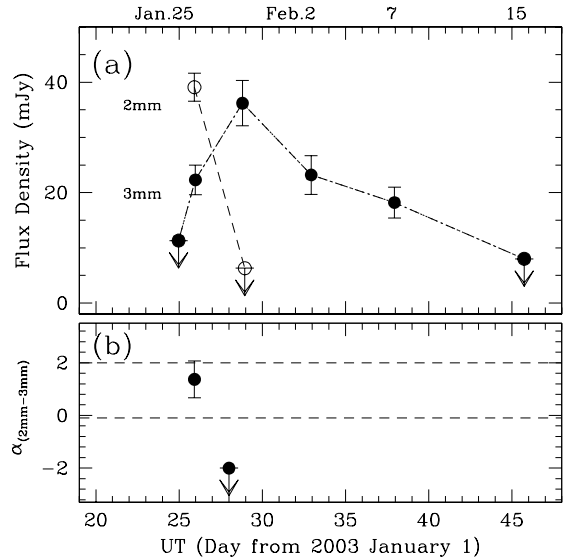}}
\caption{Time variations of 
(a) the mm continuum emission fluxes and 
(b) its spectral indices between 2 and 3 mm.
In the panel (a), open and filled circles indicate 2 and 3 mm fluxes,
respectively. The horizontal dashed lines in (b) indicate
the possible range of free-free emission (Rodriguez et al. 1993).
}
\label{fig:timevariab}
\end{figure}

\begin{table*}[ht]
\begin{center}
\caption{Summary of the Millimeter Continuum Emission Measurements}
\label{tbl:results}
\begin{tabular}{cccrccrcr}
\hline
        & & \multicolumn{2}{c}{3 mm}   & & \multicolumn{2}{c}{2 mm} & Spectral \\
\cline{3-4}\cline{6-7}
UT 2003 & & Frequency & \multicolumn{1}{c}{Flux} & & Frequency  & \multicolumn{1}{c}{Flux} & Index \\ 
        & & (GHz)     & \multicolumn{1}{c}{(mJy)} & & (GHz)      & \multicolumn{1}{c}{(mJy)} & \\ \hline
Jan.24  & &    97.782 & $<11.3$\footnotemark[$\ast$] & & $\cdot\cdot\cdot$ & $\cdot\cdot\cdot$ & $\cdot\cdot\cdot$ \\
Jan.25  & &    97.782 &   22.3$\pm$5.4 & & 146.969           & 39.1$\pm$5.1      & $+1.39\pm 0.78$ \\
Jan.28  & &    98.243 &   36.2$\pm$4.1 & & 141.135           & $<6.3$\footnotemark[$\ast$] & $<-2.4$\footnotemark[$\dagger$] \\
Feb.02  & &    99.023 &   23.3$\pm$3.5 & & $\cdot\cdot\cdot$ & $\cdot\cdot\cdot$ & $\cdot\cdot\cdot$ \\
Feb.07  & &    99.023 &   18.2$\pm$2.8 & & $\cdot\cdot\cdot$ & $\cdot\cdot\cdot$ & $\cdot\cdot\cdot$ \\
Feb.15  & &    99.023 &  $<8.0$\footnotemark[$\ast$] & & $\cdot\cdot\cdot$ & $\cdot\cdot\cdot$ & $\cdot\cdot\cdot$ \\
\hline
\end{tabular}
\end{center}
 $\ast$ $3\sigma$ upper limit, $\dagger$ 5$\sigma$ upper limit.
\end{table*}

Table \ref{tbl:results} summarizes flux densities measured 
at 2 and 3 mm integrated over one day.
We present radio ``light curves'' in Fig.\ref{fig:timevariab}a
where a clear pattern of a flare can be seen at 3 mm:~
the flux density at 3 mm increased from January 25 to 28, 
then it gradually decayed and was not detected on February 15.
Although BPB03 reported that the onset of the burst was January 20, 
we did not detect any emission at 98 GHz on January 24.
BPB03 also reported non-detections in their VLBA observations 
on January 24 at 15 and 22 GHz.
We suggest that these non-detections could be due to 
intrinsic short time variations of the source 
($\S\ref{ss:variab}$) or a possibility that 
the January 25 detection is attributed to a different 
(possibly the second giant) burst event as BPB03 suggested.
It is also possible that January 28, February 2, and 7
measurements refer to independent flares that 
may occur frequently on the star.
The behavior at 2 mm, on the other hand, was significantly different:~
though it showed bright emission on January 25.55, 
it was not detected at January 28.58.

The above time variation leads to a drastic change in the apparent spectral 
indices. We derived indices between 2 and 3 mm assuming 
a power-law type spectrum ($S_{\nu}\propto \nu^{\alpha}$):
the index, $\alpha$, observed on January 25 was positive with $+1.39\pm 0.78$, 
but the January 28 index was negative with a
5$\sigma$ upper limit of $-2.4$.
The former could be explained with a combination of
optically thin free-free emission and thermal emission emanating
from dust, but the latter cannot be easily explained.
Rodriguez et al. (1993) reported that a negative
spectral index less than $-0.1$ cannot occur for thermal processes
involving free-free emission and absorption
(Fig.\ref{fig:timevariab}b), regardless of the
electron density and temperature distribution.
In fact, we found that the neither of the indices can be
reproduced with the model of an H\emissiontype{II} region embedded 
in a dust cloud core by Furuya et al. (2002).
Cool dust radiating at those wavelengths should not exhibit 
such rapid variations of $\alpha$, 
as demonstrated by our K band observations.

\begin{figure}
\centerline{\includegraphics[angle=-90,width=72mm]{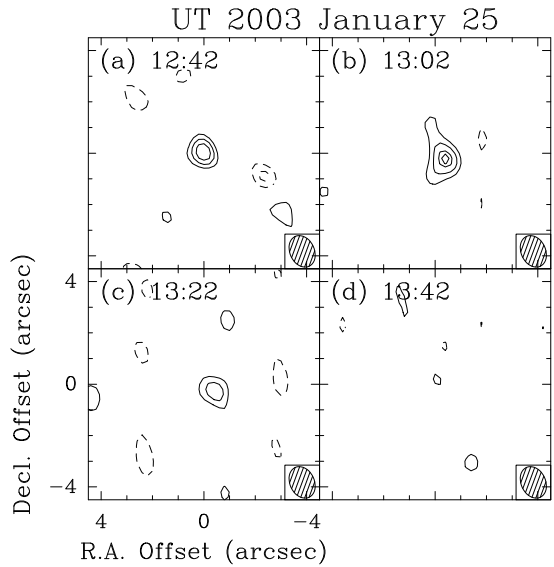}}
\caption{A series of 2 mm continuum emission maps taken  
in 2003 January 25 with a 15 min bin.
The UT shown in each panel is the center time of each integration. 
The ellipses in the bottom right corners indicate the NMA beam sizes.
Contour intervals are the same as in Fig.\ref{fig:overlay}
and the typical noise level is 12 mJy beam$^{-1}$.
}
\label{fig:4maps}
\end{figure}

\begin{figure}
\centerline{\includegraphics[angle=0,width=72mm]{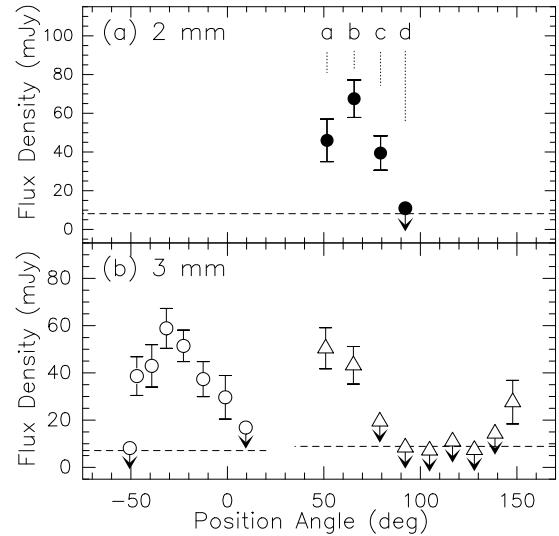}}
\caption{Plots of mm continuum fluxes vs. PAs 
(see $\S$\ref{ss:variab}) at (a) 2 mm and (b) 3 mm.
Each point is obtained with 15 min integration with an interval of 20 min. 
(a) The 2 mm data were taken on January 25:~
small labels show the corresponding panel numbers in Fig.\ref{fig:4maps}.
(b) The 3 mm data shown with open circles and triangles are
taken on January 28 and February 2, respectively.
The dotted horizontal lines show typical image noise levels.
}
\label{fig:fluxpa}
\end{figure}

\subsection{Short Time ($<$ A Day) Variations}
\label{ss:variab}

Solar and stellar flares are commonly accompanied by short 
time scale ($\lesssim$ 1 day) variations at
cm radio and X-ray wavelengths \citep{Dul85}.  
In fact, BPB03 detected short time variations at 86 GHz within 
a day of January 20.
To investigate such rapid variation, we selected 
data sets containing 15-minute (min) integrations
in which the emission was detected with signal to noise ratios 
higher than 3: 
the 2 mm data taken on January 25, and 
the 3 mm data on January 28 and February 2. 
Fig.\ref{fig:4maps} is a series of 2 mm continuum maps 
taken every 20 min on January 25.
The highly time variable nature of the source can clearly be seen.
At the beginning, its flux was 46$\pm$11 mJy (Fig.\ref{fig:4maps}a),
it flared up to 68 $\pm$ 10 mJy after 20 min (Fig.\ref{fig:4maps}b), 
then it was decaying (40$\pm$9 mJy; Fig.\ref{fig:4maps}c) and 
diminished ($3\sigma$ upper limit of 33 mJy: Fig.\ref{fig:4maps}d) 
60 min from the beginning.
During the observations,  
the calibrator was observed every 20 min with 4 min integrations, and
the amplitude of its raw visibilities taken by each baseline during 
each 4 min showed no systematic variation with fluctuation less than
12\% of the mean value.
Therefore the observed flux variations are not an artifact of
calibration. \par

It is important to discuss how linearly polarized emission 
affects the flux measurements because
radio emission showing short time variations might be 
intrinsically linearly polarized.
Receiver systems detect one direction of incident radiation.
The NMA receivers on each antenna are placed 
at the Nasmyth foci for each band and are aligned 
so that detected directions with individual antennas are 
parallel when they are projected to the plane of sky. 
Since the direction rotates while tracking the object,
observing with such systems could cause apparent flux variation.
Assuming that the radiation has a linearly polarized component 
and that its polarization angle is constant,
the phase of the flux variation should be synchronized with the
position angle (PA) of the receiver with a \timeform{180D} period.
To verify lack or existence of such a periodic pattern,
we present Fig.\ref{fig:fluxpa} where the flux variation is 
transformed as a function of PA.\par

The flux variations of the three data sets in Fig.\ref{fig:fluxpa} 
do not show any periodic pattern of phases.
The 2 mm fluxes (Fig.\ref{fig:fluxpa}a) show variation on a time
scale much shorter than a \timeform{180D} period.
This strongly suggests the 2 mm flux variation is real.
The 3 mm data taken on January 28 seem to show a considerable flux change 
peaked around PA$=$\timeform{-30D} but there is no signature
which suggests an existence of a \timeform{180D} period
(Fig.\ref{fig:fluxpa}b).	
The other 3 mm data contain too many non-detection points 
to judge whether there is a sinusoidal variation with a
\timeform{180D} period.
Moreover, we did not find a unique solution that all the 3 mm data points 
satisfy with a \timeform{180D} period,
even if we consider the overall decay of the flux (Fig.\ref{fig:timevariab}a).
We thus conclude that there is no significant contribution of
linear polarization
and 
that the observed short time variation with time scale less than 
an hour is due to intrinsic variations of the source.
These results are consistent with those from the BIMA measurements.

\subsection{The Nature of the GMR-A Flare in 2003 January}

\begin{figure}
\centerline{\includegraphics[angle=0,width=74mm]{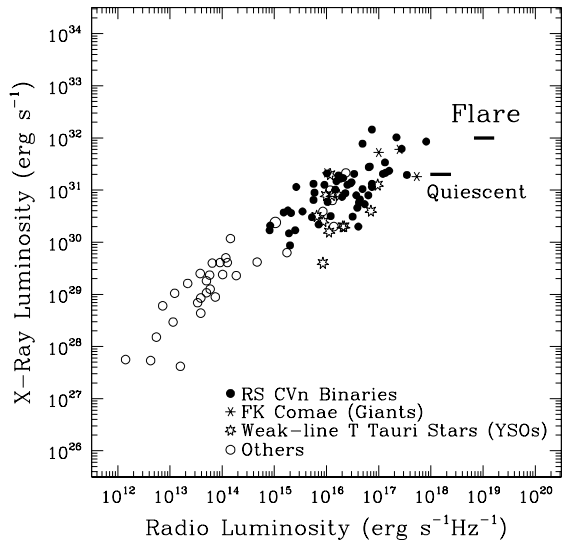}}
\caption{Comparison of $L_{\rm X}$ and $L_{\rm R}$ of GMR-A in
its quiescent and flaring phases with those for
magnetically active stars in quiescent phase
(G\"udel 2002).
The luminosity ranges for GMR-A were calculated using 
the quiescent phase cm radio fluxes reported in \citet{felli93},
quiescent and flaring phase soft X-ray fluxes in \citet{getm03} and 
the flaring phase radio fluxes shown in Fig.\ref{fig:fluxpa}.
}
\label{fig:LxLr}
\end{figure}

Fig.\ref{fig:LxLr} compares radio and soft X-ray luminosities
of GMR-A in quiescent and flaring phases with those of
magnetically active stars in quiescent phase (G\"udel 2002).  
Solar flares which are less energetic than the
stellar flares presented here are known to satisfy the correlation as well.
Considering that GMR-A is most likely to be a single WTTS (BPB03),
we conclude that the flux enhancements in 2003 January were due to 
stellar flares from a magnetically active WTTS,
such as V773 Tau (e.g., Feigelson \& Montmerle 1985).
This is because the locus of GMR-A in quiescent phase is basically 
consistent with the correlation.\par

Physical parameters for a flare can be estimated from the data of 
January 20 on which the radio and X-ray observations were carried 
out simultaneously. 
Using the soft X-ray luminosities in the flaring phase
and the time scale of the mm flare 
(BPB03; also our results from $\S\ref{ss:variab}$)
together with the reported time scale of
the X-ray flare (a few 0.1 day; Getman et al. 2003),
we estimate total energy ($E$) of the flare of $\sim 10^{35-36}$ erg,
taking $L_{\rm X}$ of $10^{32}$ erg s$^{-1}$ \citep{getm03}.
Once we know $E$,
we can pinpoint the locus of the flare in the $EM-T$ diagram 
(see Fig.~8 in [SY02]):
we estimate $T$ of $\sim 5\times 10^7$ K and $EM$ of $\sim 10^{53}$ cm$^{-3}$.
Subsequently, using Fig.2 in SY02, 
we estimate a magnetic field strength ($B$) of $50-150$ G, 
which indicates that the cyclotron frequency should appear in the GHz range.
This estimate is consistent with measurements of the stellar 
photospheric magnetic field strength (a few kG; BPB03) 
which is expected to be an order of magnitude 
stronger than those of coronal magnetic field strengths.\par

We believe that the $T$ and $B$ estimated 
above are applicable to the flares detected with the NMA.
This is because the observed flux density (Table \ref{tbl:results}) 
and the flare's size scale that can be estimated from plausible 
Alfv\'en velocity (0.05$c$ assuming $B=$100 G and
typical solar flare density of $10^9$ cm$^{-3}$) and
the flaring time scale give us the
brightness temperature of $\sim 10^8$ K, 
which is consistent with the $T$ estimated above.
In conclusion, our results are an important step toward
understanding magnetically induced stellar surface activities 
in T Tauri stars and will add to the growing evidence that 
magnetic reconnection processes could explain solar and stellar
flares over various energy scales.

\bigskip

The authors thank all the staff at Nobeyama and Okayama observatories.
Special thanks are due to Prof. K. Shibata for fruitful discussion.
The authors gratefully acknowledge the anonymous referee whose comments
significantly improved the quality of the paper.
We wish to acknowledge discussions with
R. Bandiera, 
R. Cesaroni,
M. Felli,
M. G\"udel,
K. Hamaguchi,
H. Hirashita,
S. Kameno,
K. Shibasaki,
M. Shimojo,
S. Takahashi,
L. Testi,
C. M. Walmsley,
A. Wootten, and
K. Yanagisawa.


\begin{thebibliography}{}

\bibitem[Bower, Plambeck, \& Bolatto(2003b)]{Bow03a} Bower, 
G.~C., Plambeck, R., \& Bolatto, A.\ 2003a, \iaucirc, 8055, 2

\bibitem[Bower et al.(2003b)]{Bow03b} Bower, 
G.~C., Plambeck, R., Bolatto, A., McGrady, N., Graham, J. R.,
de Pater, I., Lin., M. C., \& Baganoff, F. K. \ 2003b, \apj, in press (BPB03)

\bibitem[Dulk (1985)]{Dul85} Dulk, G.~A.\ 1985, \araa, 23, 169 

\bibitem[Feigelson \& Montmerle(1985)]{feig85} Feigelson, E.~D.,~\& Montmerle, T.\ 1985, \apjl, 289, L19 

\bibitem[Feldman, Laming \& Doschek (1995)]{feld95} Feldman, 
U., Laming, J.~M., \& Doschek, G.~A.\ 1995, \apjl, 451, L79 

\bibitem[Felli et al.(1993)]{felli93} Felli, M., Taylor, G.~B., 
Catarzi, M., Churchwell, E., \& Kurtz, S.\ 1993, \aaps, 101, 127 

\bibitem[Furuya et al.(2002)]{Furu02} Furuya, R.~S., Cesaroni, 
R., Codella, C., Testi, L., Bachiller, R., \& Tafalla, M.\ 2002, \aap, 390, L1 

\bibitem[Garay, Moran, \& Reid(1987)]{Gar87} Garay, G., 
Moran, J.~M., \& Reid, M.~J.\ 1987, \apj, 314, 535

\bibitem[Getman et al.(2003)]{getm03} Getman, K.~V., 
Feigelson, E.~D., Garmire, G., Murray, S.~S., \& Harnden, F.~R.\ 2003, 
\iaucirc, 8068, 2 

\bibitem[G{\" u}del(2002)]{guedel02} G{\" u}del, M.\ 2002, \araa, 40, 217 

\bibitem[Hillenbrand \& Carpenter(2000)]{Hill00} Hillenbrand, L.~A.,~\& Carpenter, J.~M.\ 2000, \apj, 540, 236 

\bibitem[Muench, Lada, Lada, \& Alves(2002)]{Mue02}  Muench, A.~A., Lada, E.~A., Lada, C.~J., \& Alves, J.\ 2002, \apj, 573, 366 

\bibitem[Nakanishi et al.(2003)]{Nakanishi03} Nakanishi, K., Saito, M., Furuya, R.~S., Shinnaga, H., \& Momose, M.\ 2003, \iaucirc, 8060, 2 

\bibitem[Rodriguez et al.(1993)]{Rod93} Rodr\'iguez, L.~F., 
Marti, J., Canto, J., Moran, J.~M., \& Curiel, S.\ 1993, Rev. Mex. Astron. Astrofis., 25, 23 

\bibitem[Shibata \& Yokoyama(1999)]{Shibata99} Shibata, K.~\& Yokoyama, T.\ 1999, \apjl, 526, L49

\bibitem[Shibata \& Yokoyama (2002)]{Shibata02} Shibata, K.~\& Yokoyama, T.\ 2002, \apj, 577, 422 (SY02)


\end{thebibliography}
\end{document}